\documentclass[journal]{IEEEtran}
\ifCLASSINFOpdf
\else
\fi
\usepackage{amsbsy}
\usepackage{amsmath}
\usepackage{amsfonts}
\usepackage{graphicx}
\usepackage{epsfig}
\usepackage{multirow}
\usepackage{array}
\newcolumntype{L}[1]{>{\raggedright\let\newline\\\arraybackslash\hspace{0pt}}m{#1}}
\newcolumntype{C}[1]{>{\centering\let\newline\\\arraybackslash\hspace{0pt}}m{#1}}
\newcolumntype{R}[1]{>{\raggedleft\let\newline\\\arraybackslash\hspace{0pt}}m{#1}}

\usepackage{scrextend}
\usepackage{hanging}
\usepackage{flushend}

\begin{document}
\title{Risk Constrained Trading Strategies for Stochastic Generation with a Single-Price Balancing Market}

\author{Jethro~Browell
\thanks{J. Browell is with the Department
of Electronic and Electrical Engineering, University of Strathclyde,
Glasgow, UK, e-mail: jethro.browell@strath.ac.uk.}% <-this % stops a space
\thanks{Manuscript Submitted \today}}

\maketitle

\begin{abstract}
Due to the limited predictability of wind power and other stochastic generation, trading this energy in competitive electricity markets is challenging. This paper derives revenue-maximising and risk-constrained strategies for stochastic generators participating in electricity markets with a single-price balancing mechanism. Starting from the optimal---and impractical---strategy of offering zero or nominal power, which exposes the participant to potentially large imbalance costs, we develop a number of strategies that control risk by hedging against penalising balancing prices in favour of rewarding ones. Trading strategies are formulated in a probabilistic framework in order to address asymmetry in balancing prices. The large-scale communication of system information characteristic of modern power systems is utilised to inputs for electricity price forecasts and probabilistic system length forecasts. A case study using data from the GB market in the UK is presented and the ability of the proposed strategies to increase revenue and reduce risk is demonstrated and analysed.
\end{abstract}


% \fi
\IEEEpeerreviewmaketitle

\section{Introduction}
%\IEEEPARstart{T}{his} demo file is intended to serve as a ``starter file''
%for IEEE journal papers produced under \LaTeX\ using
%IEEEtran.cls version 1.8a and later.

%\begin{table}[h]
%%	\caption{List of Symbols}
%	\subsection{List of Symbols}
%	\begin{tabular}{lL{10cm}}
%		$R_{t+k}$ & Revenue \\
%		$E^\text{C}_{t+k}$ & Contracted Volume in Day-ahead Auction\\
%		$E_{t+k}$ & Actual Energy Delivered \\
%		$E_\text{max}$ & Nominal Deliverable Volume of Participant \\
%		$\pi^\text{C}_{t+k}$ & Day-ahead Auction Price \\
%		$\pi^\text{S}_{t+k}$ & Main Price \\
%		$\pi^+_{t+k}$ & Main Price (Net Up-regulation) \\
%		$\pi^-_{t+k}$ & Main Price (Net Down-regulation)\\
%		$T_{t+k}$ & Imbalance Cost \\
%		$\hat{x}$ & Forecast or Estimate of $x$ \\
%		$\cal{E}\left\{\cdot\right\}$ & Expectation Operator \\
%	\end{tabular}
%\end{table}

% Why are we interested in these offer strategies?
\IEEEPARstart{E}{lectricity} markets were designed for dispatchable generation. Since energy liberalisation in Europe, the US, and elsewhere, supply and demand have been matched by centralised operation of transmission systems using a combination of connected markets for energy and ancillary services~\cite{Edwards2009}. Stochastic generators, such as wind and solar, require power production forecasts to participate economically in these markets, and while high-quality forecasts are widely available and improving, they will never be perfect. 

Much attention has been given to how stochastic generation can be integrated into electricity markets~\cite{Morales2014}. Strategies for their participation can benefit from use of information about forecast uncertainty in order to deal with asymmetric penalties for over- or under-producing. Most of this work has focused on wind since this technology is more established, but the principals are transferable to other stochastic generators, such as solar, or other smart grid actors such as aggregators~\cite{Pei2016}. Looking towards the future, new electricity market arrangements which facilitate the active participation of renewables in balancing markets are becoming a reality~\cite{Fernandes2016}, and market designs that incorporate directly properties of stochastic generation by allowing probabilistic offers have been proposed~\cite{Papakonstantinou2016}.

Offer strategies for trading wind power in day-ahead markets are derived in~\cite{Bathurst2002}, and for dual-price balancing markets the optimal quantile of a predictive distribution can be calculated based on forecasts of imbalance prices~\cite{Bremnes2004,Pinson2007}. This analysis has been extended in~\cite{Bitar2012} to include a recourse offer closer to the time of delivery. Participation in an intraday power exchange (which facilitates anonymous bilateral trading) is considered in~\cite{Skajaa2015} by accepting available bids and offers which are deemed improve the participant's market position, though without a sophisticated offer strategy for the day-ahead market. The strategic behaviour of wind power as a price-maker has also been studied~\cite{Zugno2013,Delikaraoglou2015}.

Much of this work has been presented from the perspective of the power forecaster; however, participants in electricity markets also require price forecast to inform their decision making, as the examples given above acknowledge. Electric utilities forecast prices from hours to months ahead in order to reduce risk or maximise profits, with day-ahead and intraday forecasts critical for effective bidding strategies~\cite{Edwards2009}. An extensive review of electricity price forecasting can be found in~\cite{Weron2014} and results of the price forecasting track of the 2014 Global Energy Forecasting Competition in~\cite{Hong2016} give an overview of state-of-the-art practices. Electricity prices are driven fundamentally by supply and demand costs, and as such it is necessary to model these when making predictions~\cite{Karakatsani2008,Brijs2015,Ziel2015}. Much attention has been given to short-term electricity prices, with familiar time-series models (ARX, ARIMA, etc) being popular. Forecasting balancing prices has received less attention but is considered in~\cite{Olsson2008} and balancing volumes in~\cite{jonsson2014}, who use approaches based on ARIMA and exponential smoothing, respectively.

While a great deal has be learned about how stochastic generators
can participate in electricity markets the majority of this work has focused on markets a with dual-price balancing mechanism. This is largely due to the high penetration of renewables in these markets, particularly in Europe. While the majority of electricity markets in Europe operate dual-pricing systems, single-price markets dominate in the US and are operated in Germany, the Netherlands and, since November 2015, the UK.

Large volumes of system data that are collected and shared in smart grids to enable efficient use of available assets and resources~\cite{Fang2012}. This data may also be utilised in the strategies of electricity market participants. Furthermore, since this information is available to participants electronically it can easily be incorporated into automated trading systems.

In this work, participation of stochastic generation in a day-ahead energy market coupled with a single-price balancing balancing market is considered. It is observed that in this situation, electricity market forecasts are of primary importance and that power forecasts are required only to apply risk-constraints. A secondary result is the observation of weak incentives for variable generation to provide accurate forecasts in this scenario, as also observed in~\cite{Hirth2015}. Strategies based on taking a long or short position in order to manage asymmetric imbalance costs are proposed. Probabilistic forecasts of system length (sign of the net system imbalance) are required in addition to forecasts of day-ahead and balancing prices. Revenue maximisation and risk-constrained strategies are both derived, with only the latter requiring forecasts of power production.

An introduction to day-ahead and balancing markets is offered in Section~\ref{sec:MarketsIntro} followed by the formulation of the offer strategy problem and possible solutions in Section~\ref{sec:problemFormulation}. Probabilistic system length forecasts using logistic regression, and Price forecast using ARMAX models, both of which utilise wider system data, are described in~\ref{sec:SLandPriceForecasting}. A case study using real data form the UK is presented in Section~\ref{sec:CaseStudy}. Forecast performance is compared to standard benchmarks and quantified in monetary terms based on the performance of trading strategies. Finally, concluding remarks are made in Section~\ref{sec:Discussion}.

\section{Day-ahead and Balancing Markets}
\label{sec:MarketsIntro}

% How do electricity markets work?
Market structures can vary widely between
regions but are typically made up of four main components:
bi-lateral contracting between individual parties from
days to years ahead of delivery, a day-ahead auction that determines the schedule for the activation of generators and large industrial consumers for the following day, an intraday market which allows participants to modify their position closer to delivery, and a balancing market utilised by the transmission system operator to balance supply and demand in real time~\cite{Harris2006}. Additional markets may also exist for ancillary services such as frequency response, reserve power and provision of reactive power, and financial products such as energy options and futures.

The most important markets for stochastic generators are the day-ahead and the balancing markets since the need to forecast generation makes trading further in advance impractical and because  forecast errors result in imbalances. Intraday markets enable participants to modify their position closer to gate closure, but often suffer from low liquidity meaning that it is difficult for participants to find counter-parties to trade with. 

Day-ahead markets are typically double-blind auctions into which generators and consumers submit anonymous offers to generate and bids to consume certain volumes of energy at a price they are willing to pay or be paid. Supply and demand are compared and a market price is calculated for each period of the following day. This price is applied to all accepted bids and offers and is an important reference for intraday and balancing markets since it gives an indication of the marginal price of energy for a given period.

% What's the deal with balancing?
% Balancing markets are of particular concern to stochastic generators since, as a result of imperfect forecasts, they are almost certain to be `out of balance' most of the time.
Balancing markets are used by the transmission system operator to balance supply and demand and operate from from gate closure to the point of delivery. The cost of balancing incurred by the system operator is recovered through payments by those who are out of balance. Prices are calculated based on either a single- or dual-pricing system. All participants in a single-price system resolve their imbalance at the same imbalance price, whereas in a dual-price system participants receive different prices depending on the sign of their imbalance. Balancing prices represent the cost to the system operator of increasing or decreasing net-generation, and as such depend on whether the system has a net energy surplus or deficit.

In a single-price balancing market, whether the single price
is greater or less than the day-ahead price depend on the system length, i.e. whether the transmission system operator has had to increase or decrease net-energy production during a given time period. If the system is short of energy, the balancing price will be greater than the day-ahead price the reflect the utilisation of more expensive or flexible generators (or demand reduction), and the converse if the system is long. The effect of this is to penalise market participants
who are out-of-balance in the same direction as the system, and to reward those who are helping the system by being out-of-balance in the opposite direction. This is different to the two-price system where imbalances contributing to the system imbalance are penalised, and those helping receive a neutral reference price, which is usually similar to the day-ahead price.

The importance of system length forecasting is clear: being out-of-balance
the \textit{wrong} way invites a penalty, whereas being out-of-balance the \textit{right} way is profitable. However, forecasting the system length prior to submitting offers into the day-ahead market is challenging and, since the penalties and rewards for a correct forecast are not systemic, warrants a probabilistic approach. In the following section this problem is formulated and offer strategies for the day-ahead market are derived based on a probabilistic assessment of system length.

\section{Problem Formulation}
\label{sec:problemFormulation}

Here we consider day-ahead offer strategies for a participant who
is a price-taker in both day-ahead and balancing markets, and do not consider participation in intraday markets.
For each settlement period $t+k$, a market participant will contract some volume of energy $E^\text{C}_{t+k}$ at time $t$. The revenue $R_{t+k}$ for a participant contracting $E^\text{C}_{t+k}$ but generating $E_{t+k}$ is given by
\begin{equation}
	R_{t+k} = \pi^\text{C}_{t+k} E^\text{C}_{t+k} - T^\text{C}_{t+k}
\end{equation}
where $\pi^\text{C}_{t+k}$ is the contracted price for period $t+k$,
and $T^\text{C}_{t+k}$ is the cost associated with the energy imbalance
$d_{t+k} =  E^\text{C}_{t+k} - E_{t+k}$.
In a single-price balancing market, each participant must buy the volume of energy equal to their
deficit, or sell the volume equal to their surplus, at the imbalance price.
The imbalance price $\pi^\text{S}_{t+k}$ is a function of the balancing actions relating to period $t+k$ taken by the system operator to maintain the balance supply and demand in real time and is calculated at the each settlement period.
The imbalance cost $T^\text{C}_{t+k}$ is given by
\begin{equation}
	T^\text{C}_{t+k} = \pi^\text{S}_{t+k} d_{t+k} \quad .
\end{equation}
It is useful to express a market participant's revenue in terms of the actual energy they generate and their imbalance as follows
\begin{equation}
	R_{t+k} = \pi^\text{C}_{t+k} E_{t+k} - T_{t+k} 
\end{equation}
from which it is clear that in order to maximise revenue, balancing costs
\begin{equation}
	T_{t+k} = \left(\pi^\text{S}_{t+k} - \pi^\text{C}_{t+k} \right)d_{t+k} \quad ,
	\label{eqn:balCosts}
\end{equation}
should be minimised.

%Unlike a dual-price balancing market, the same price is applied to
%imbalances regardless of sign of the participant's imbalance. However,
%this price reflects the cost of balancing the system as a whole
%and has a dual-nature due to the different marginal cost of up- and
%down-regulation. In general, if the system is long, i.e. has excess energy,
%the imbalance price will be lower than the contracted price and therefore
%penalise market participants who are also long, since they could have got
%a better price fore their excess energy in the day-ahead market. This
%situation also rewards those who are short, since they pay less to make
%up their deficit than they were paid for the energy they contracted for
%but didn't generate.
%The converse is also true: if the system is short,
%the imbalance price is greater than the contracted price, penalising those
%who have contributed to the system imbalance and rewarding those who have not.

In order to reflect the dual-nature of the single imbalance price,
we distinguish between the price resulting from net up- or down-regulation, which corresponds to the sign of the system net imbalance volume (NIV).
Equation~\eqref{eqn:balCosts} then becomes
\begin{equation}
	T_{t+k} =
	  \begin{cases}
	    \left(\pi^+_{t+k} - \pi^\text{C}_{t+k} \right)d_{t+k}  & \quad \text{if} \quad \text{NIV}_{t+k} > 0 \\
	   % \hfil 0  & \quad \text{if} \quad \text{NIV}_{t+k} = 0 \\
	    \left(\pi^-_{t+k} - \pi^\text{C}_{t+k} \right)d_{t+k}  & \quad \text{if} \quad \text{NIV}_{t+k} \le 0 \\ 
	  \end{cases} \quad .
	  \label{eqn:updownbalcosts}
\end{equation}
where $\pi^+_{t+k}>\pi^\text{C}_{t+k}$ is the up-regulation price, and $\pi^-_{t+k}<\pi^\text{C}_{t+k}$ is the down-regulation price. The case $\text{NIV}_{t+k} = 0$ is merged with $\text{NIV}_{t+k}<0$ for simplicity and without loss of generality since $\pi^+_{t+k}=\pi^-_{t+k}=\pi^\text{C}_{t+k}$ in that situation.

Assuming not participation in other markets or incentives to do otherwise, the aim of the market participant is to contract the volume of energy $E^\text{C}_{t+k}$ that maximises revenue while managing risk.

\subsection{Imbalance Minimisation}
\label{sec:p50}

First we consider the simplest strategy for risk management: minimise exposure to imbalance charges by contracting the forecast generation for each period in the day-ahead market. This approach reduces exposure to penalising imbalance prices, but also reduces exposure to rewarding prices in the case where the sign of the participant's imbalance is the opposite that of the system.
The bid in this case is given by
\begin{equation}
	E^\text{C}_{t+k} = \hat{E}_{t+k|t} \quad ,
\end{equation}
where $\hat{E}_{t+k|t}$ is a forecast of $E_{t+k}$ made at time $t$ set to minimise the mean absolute error. This strategy has the benefit of not requiring forecasts of market prices or system length.

\subsection{Categorical Assessment of System Length}
\label{sec:detLength}

If the system length were known at the time of contracting, according to Equation~\eqref{eqn:updownbalcosts}, the optimal volume to contract would be $\pm\infty$! This is of course nonsense and in violation of the price-taker assumption since such offers would influence the clearing price of the day-ahead market and NIV.

Any participant with sufficient power would have to conciser their influence on the day-ahead price, NIV and the marginal price of balancing actions that the system operator would have to take, and the opinion of the market regulator. This situation is not considered here. We therefore proceed assuming that the capacity of the wind generators we consider is small relative to the magnitude of the NIV, and that the contracted volume is restricted to the range $0\le E^\text{C}_{t+k} \le E_\text{max}$, where $E_\text{max}$ is maximum amount of energy the wind generator could deliver in a single settlement period.

If the sign of the NIV for period $t+k$ is known, the optimal bid would be simply
\begin{equation}
	E^\text{C}_{t+k} =
	  \begin{cases}
	    0  & \quad \text{if} \quad \text{NIV}_{t+k} > 0 \\
	   % E_{t+k}  & \quad \text{if} \quad \text{NIV}_{t+k} = 0 \\
	    E_\text{max}  & \quad \text{if} \quad \text{NIV}_{t+k} \le 0 \\ 
	  \end{cases} \quad .
\end{equation}
A deterministic forecast of the sign of the NIV is required to implement this strategy, but no power forecast is needed. Note also that bidding only extremes leaves the participant exposed to potentially large losses if the sign of the NIV is forecast incorrectly.

\subsection{Probabilistic Assessment of System Length}
\label{sec:probLength}

As imbalance prices in periods of net up and down regulation are asymmetric about $\pi^\text{C}$, it is desirable to formulate offer strategies from a  probabilist perspective. Consider the energy generated during period $t+k$ to be a random variable $E_{t+k}$, and the probability at time $t$ that the system will be short $\Pr_t(\text{NIV}_{t+k}>0)=\phi_{t+k|t}$.
We include the possibility that the NIV is exactly zero in the chance that the
system is long without consequence, so $\Pr_t(\text{NIV}_{t+k}\le 0)=1-\phi_{t+k|t}$.

The subscripts $t+k$ and $t+k|t$ are dropped in the proceeding analysis to avoid notational clutter.

In this probabilistic framework the expectation of the imbalance cost $T$
is given by
\begin{equation}
	T = \phi\left(\pi^+ - \pi^\text{C} \right)d + (1-\phi)\left(\pi^- - \pi^\text{C} \right)d
\end{equation}
where $d=E^\text{C}-E$. Using the expectation operator ${\cal E}\{\cdot\}$, the optimal bid can now be calculated as
\begin{eqnarray}
	\tilde{E}^\text{C} &=& \underset{E^\text{C}}{\operatorname{argmin}}~{\cal E}\left\{ T \right\} \\
	&=&  \underset{E^\text{C}}{\operatorname{argmin}}~
		\left(\phi(\pi^+ - \pi^-) + \pi^- - \pi^\text{C} \right)
		{\cal E}\left\{
			d
		\right\} \\
	&=&  \underset{E^\text{C}}{\operatorname{argmin}}~
		\Big[\left(\phi(\pi^+ - \pi^-) + \pi^- - \pi^\text{C} \right) \notag \\
        && \hspace{3cm} \times
		\left(E^\text{C}-{\cal E}\left\{
					E
				\right\}\right) \Big]
\end{eqnarray}
Since the optimal bid depends only on the sign of the factor
multiplying the expected imbalance, it is helpful to define the
ratio
\begin{equation}
	\Phi = \frac{\pi^\text{C} - \pi^-}{\pi^+ - \pi^-}
	\label{eqn:critProb}
\end{equation}
and write the optimal bid as
\begin{equation}
	\tilde{E}^\text{C} =
	\begin{cases}
		E_\text{max} & ~ \text{if} \quad \phi<\Phi	\\
		0			& ~ \text{if} \quad \phi \ge \Phi
	\end{cases} \quad .
\end{equation}
The ratio \eqref{eqn:critProb} can be interpreted as a cost/loss ratio
defining the critical probability at which is becomes economic to bid
as if the system is expected to be long or short.

As the prices $\pi^\text{C},~\pi^+$ and $\pi^-$ are unknown at time $t$ they must be forecast along with $\phi$. Note, however, that there is no need to forecast the level of wind generation.

%%%%%%%%%%%%%%
% \begin{equation}
% 	E^\text{C} =
% 	\begin{cases}
% 		E_\text{max} & ~ \text{if} \quad \phi<\frac{\pi^\text{C} - \pi^-}{\pi^+ - \pi^-}	\\
% 		0			& ~ \text{if} \quad \phi \ge \frac{\pi^\text{C} - \pi^-}{\pi^+ - \pi^-}
% 	\end{cases} \quad .
% \end{equation}
%%%%%%%%%%%%%%

\subsection{Risk Constrained Contracted Volume}
\label{sec:risk_constrained}

So far we have only considered revenue maximisation. Next we consider a risk constrained approach for two main reasons: first,
the risk associated with revenue maximisation is potentially large
since imbalance prices are volatile and the strategies investigated so far require the participant to expose themselves to the largest imbalance possible; and second, participants with potential market power may be able to participate in a similar way by hedging smaller volumes, and this should be done in an informed way.

In this section alternative strategies are considered that restrict
the size of the expected imbalance by adjusting the offer away from the forecast generation $\hat{E}={\cal E}\{E\}$
in order to hedge against penalising imbalance prices.
% Three options are considered: an additive adjustment where the offer is equal to expected generation plus/minus some parameter, $\nu$, a multiplicative adjustment where the offer is equal to expected generation multiplied by some factor, $1\pm \eta$, and a variable adjustment based on limiting expected positive imbalance charges.
Three options are considered: an additive adjustment where the offer is equal to expected generation plus/minus some parameter, $\nu$; a multiplicative adjustment where the offer is equal to expected generation multiplied by some factor, $1\pm \eta$;and finally, offering a quantiles of a probabilistic generation forecasts.

\subsubsection{Additive Adjustment}

In this strategy the contracted energy for a given settlement period is the expected energy plus/minus a fixed adjustment. In effect, the capacity is partitioned into $\hat{E}-\nu E_\text{max}$ and $2\nu E_\text{max}$ with the latter part traded using the probabilistic forecast of system length. The final offer bound by $0\le E^\text{C} \le E_\text{max}$. This strategy can be written as
\begin{equation}
	E^\text{C} =
		\begin{cases}
			\min\big\{ E_\text{max} ,~\hat{E}+\nu E_\text{max} \big\} & ~ \text{if} \quad \phi<\Phi	\\
			\max \big\{ 0 ,~\hat{E}-\nu E_\text{max} \big\}			& ~ \text{if} \quad \phi \ge \Phi
		\end{cases} \quad .
		\label{eqn:constOffer1}
\end{equation}
The choice of $\nu$ is a trade-off between maximising revenue and reducing exposure to imbalance charges.

% Using equation~\eqref{eqn:balCosts}, the imbalance a single settlement period for this offer strategy can be written
% \begin{equation}
% 	T =  \left(\pi^\text{S} - \pi^\text{C} \right) \left(\hat{E}+\nu E_\text{max}-E \right)
% \end{equation}

\subsubsection{Multiplicative Adjustment} 

Here we consider a contracted volume proportional to the
forecast generation. This strategy has the pleasing property that
exposure to imbalance charges increases with expected generation,
and therefore with expected revenue for a given period. Put differently, the participant is only exposed to risk when the expected revenue is already high, and is exposed to little risk when expected revenue is low. The contracted volume is equal to $\hat{E} \pm (\eta \times 100)\%$,
bound by zero and $E_\text{max}$. The strategy given by
\begin{equation}
	E^\text{C} =
		\begin{cases}
			\min \big\{ E_\text{max},~(1+\eta)\hat{E} \big\} & ~ \text{if} \quad \phi<\Phi	\\
			\max \big\{0,~(1-\eta)\hat{E} \big\}		& ~ \text{if} \quad \phi \ge \Phi
		\end{cases}
		\label{eqn:propOffer1}
\end{equation}
where $\eta \ge 0$.

\subsection{Quantile Offer}

The additive and multiplicative strategies result in an imbalance $d$ equal to the wind power forecast error, $\hat{E}-E$, plus or minus an adjustment, the aim being to increase the likelihood that this term is either positive of negative, depending on the values of $\phi$ and $\Phi$. Probabilistic forecasts provide information about uncertainty associated with forecast errors. This information can be used to chose $E^\text{C}$ such that the probability of $d>0$ is a specific value.

The predictive distribution of $E$ can be described by a set of quantiles $\{q_\alpha,\alpha \in [0,1]\}$ where
\begin{equation}
	\Pr \left( E < q_\alpha \right) = \alpha \quad .
\end{equation}
Writing this in terms of $d$ and $E^\text{C}$ gives $\Pr \left( d < q_\alpha - E^\text{C} \right) = \alpha$. Therefore, the contracted volume $E^\text{C}$ which results in a probability $\alpha$ of $d$ being negative is given by the quantile $q_\alpha$. This strategy is written as
\begin{equation}
	E^\text{C} =
		\begin{cases}
			q_{\alpha^\prime} & ~ \text{if} \quad \phi<\Phi	\\
			q_{1-\alpha^\prime} & ~ \text{if} \quad \phi \ge \Phi
		\end{cases}
		\label{eqn:quantOffer}
\end{equation}
where $\alpha^\prime$ is the probability that the realisation of $d$ has the desired sign.

This approach is attractive because it explicitly models the uncertainty associated with forecast errors allowing this risk-factor to be controlled explicitly. It is also more elegant since it removes the need to impose bounds on offers as quantiles are bound by $[0,E_\text{max}]$ automatically.

% \subsubsection{Variable Adjustment}

% In order to control exposure to risk for specific settlement periods, we calculate an adjustment
% for each period based on the expectation of the positive imbalance cost, which is given by
% \begin{equation}
% 	T^\text{R} =
% %	r =
% 		\begin{cases}
% 			\phi(\pi^+ - \pi^\text{C})(E^\text{C}-\hat{E}) & ~ \text{if} \quad \phi<\Phi \\ %~\text{and}~\eqref{eqn:constOffer1}	\\ 
% 			(1-\phi)(\pi^\text{C} - \pi^-)(E^\text{C}-\hat{E}) & ~ \text{if} \quad \phi \ge \Phi %~\text{and}~\eqref{eqn:constOffer1}
% 		\end{cases} \quad .
% \end{equation}
% For some maximum expected imbalance cost, $r$, the value of $\nu=\nu_{t+k}$ or $\eta=\eta_{t+k}$ is maximised subject to $T^\text{R} \le r$. In this case the both offer strategies~\eqref{eqn:constOffer1} and~\eqref{eqn:propOffer1} become,
% \begin{equation}
% 	E^\text{C} =
% 		\begin{cases}
% 			\min\left\{ E_\text{max} ,~\hat{E}+\dfrac{r}{\phi(\pi^+ - \pi^\text{C})} \right\} & ~ \text{if} \quad \phi<\Phi	\\
% 			\max \left\{ 0 ,~\hat{E}- \dfrac{r}{(1-\phi)(\pi^\text{C} - \pi^-)}\right\}			& ~ \text{if} \quad \phi \ge \Phi
% 		\end{cases} \quad .
% 		\label{eqn:Offer2}
% \end{equation}
% Note that $r$ is not simply a measure of risk: in order to
% minimise positive imbalance costs it will be necessary for $r$ (or $\nu$) to be positive
% for hedging, and therefore risk reduction, to take place.

\section{Forecasting}
\label{sec:SLandPriceForecasting}

\subsection{Probabilistic System Length Forecast}

The probability that the system is will be short, $\phi$, is estimated using a logistic regression model. This approach allows $\phi$ to be estimated conditional on some set of explanatory variables $\mathbf{X}$, formally,
\begin{equation}
	\phi = \Pr(\text{NIV}>0 | \mathbf{X}) \quad .
\end{equation}
The logistic regression model is given by
\begin{equation}
	\log \frac{\phi}{1-\phi} = \boldsymbol\beta \cdot \mathbf{X}_{t+k}
\end{equation}
where the vector $\boldsymbol{\beta}$ contains the model parameters to be estimated. Solving for $\phi$ yields
\begin{equation}
	\phi = \frac{1}{1+\exp({-\boldsymbol\beta \cdot \mathbf{X}_{t+k}})} \quad .
\end{equation}
Explanatory variables are chosen from the wide range of power system and market data that are available to participants. In this work, the parameters $\boldsymbol\beta$ are determined by maximum likelihood estimation using \texttt{R}, specifically the function \texttt{glm} from the package \texttt{stats}.
Deterministic system length forecasts are produced using the same method but rounding $\phi \ge 0.5$ to 1, and $\phi<0.5$ to 0.

% The logistic function has been applied in many situations to estimate the conditional probability of binary outcomes; however, it may be desirable to employ a more flexible function to improve forecast calibration. As the value of probability $\phi$ determines where the participant takes a long or shot position, calibration is critical. A generalisation of the logistic function is considered in order to attain the best forecast possible. Consider the model given  by
% \begin{equation}
% 	\log \frac{\phi^\nu}{1-\phi^\nu} = \boldsymbol\beta \cdot \mathbf{X}_{t+k}
% \end{equation}
% and
% \begin{equation}
% 	\phi = \left[\frac{1}{1+\exp({-\boldsymbol\beta \cdot \mathbf{X}_{t+k}})}\right]^\frac{1}{\nu} \quad ,
% \end{equation}
% where $\nu>0$ is the shape parameter of the generalised logistic function~\cite{Mead1965}. The shape parameter controls the relative growth of the logistic function close to the asymptotes of 0 and 1 enabling the predictor model a degree of asymmetry in the response of the predictor around the critical value of $\boldsymbol\beta \cdot \mathbf{X}_{t+k}=0$.

\subsection{Price Forecasts}

For the purpose of this study, we employ the popular ARMAX-type models for price forecasting~\cite{Weron2014}. A separate ARMAX model is fit for each settlement period and type of day to capture the different dependencies between price and exogenous variables in each situation. The time index $\tau$ is used to indicate the position of price $\pi_\tau$ in a sequence of prices corresponding to the settlement period and day-type. This approach regresses the price at time on its past values at $\tau-1$,...,$\tau-p$, the the model error $\epsilon_\tau$ and exogenous variables $X_{k,\tau}$. The model is written 
\begin{equation}
\pi_\tau = \alpha_0  + \epsilon_\tau+  \sum^p_{i=1} \alpha_i \pi_{\tau-i} + \sum^q_{j=1} \beta_j \epsilon_{\tau-j} + \sum_k \gamma_k X_{k,\tau}
\end{equation}
where $\alpha_i$ are the autoregressive coefficients, $\beta_j$ are the moving average coefficients, and $\gamma_k$ are the regression coefficients for the exogenous variables. The forecast of $\pi_\tau$ is given by
\begin{equation}
\hat{\pi}_\tau = \alpha_0  + \sum^p_{i=1} \alpha_i \pi_{\tau-i} + \sum^q_{j=1} \beta_j \epsilon_{\tau-j} + \sum_k \gamma_k X_{k,\tau} \quad .
\end{equation}
%The error, $\epsilon_\tau = \pi_\tau - \hat{\pi}_\tau$, can therefore be calculated once the value of $\pi_\tau$ is know.
The parameters $\alpha_i,~\beta_j$ and $\gamma_k$ are determined by maximum likelihood expectation and the model order $(p,q)$ by minimising the Akike Information Criterion implemented using the \texttt{R} package \texttt{forecast}~\cite{Hyndman2008}.

% For the purpose of this study, we employ a popular benchmark method for price forecasting based on exponential smoothing combined with the `similar day' method. In effect, the prices for each settlement period and type of day (weekday, weekend or holiday) are smoothed separately. For convenience, the balancing prices and day-ahead price are contained in the vector	$\boldsymbol{\Pi}_{t+k} = \left(\pi_{t+k}^+,~\pi_{t+k}^\text{C},~\pi_{t+k}^-\right)^\text{T}$. The price forecasts made at  time $t$ for period $t+k$ is given by
% \begin{equation}
% 	%\hat{\pi}_{t+k} = \hat{\pi}_{(t,k,D)} = \lambda \hat{\pi}_{(t-T,k,D)} + (1-\lambda)\pi_{(t-T,k,D)}
%     %\hat{\pi}_{t+k} = \lambda \hat{\pi}_{t+k-T_D} + (1-\lambda)\pi_{t+k-T_D}
%     \hat{\boldsymbol \Pi}_{t+k}=  \lambda \hat{\boldsymbol \Pi}_{t+k-T_D} + (1-\lambda){\boldsymbol \Pi}_{t+k-T_D}
% \end{equation}
% where $T_D$ is the length of time between $t$ and the previous day of the same type, and $\lambda$ is the smoothing parameter. 

\subsection{Wind Power Quantile Forecasts}

Quantile forecasts for time $t$, $\hat{q}_{\alpha,t}$ are given by the function, $\hat{q}_{\alpha,t}=Q_\alpha(\theta_t)$, of explanatory variables, $\theta_t$, that is the solution to the following optimisation problem
\begin{equation}
% \underset{q}{\operatorname{argmin}} \left[ 
% 		(1-\alpha)\sum_{y_i<q}(q-y_i) + \alpha\sum_{y_i\ge q}(y_i-q)
% 	\right] \quad .
\underset{Q_\alpha}{\operatorname{argmin}} \left[ 
		\sum_t\max \left\{
        (1-\alpha)(\hat{q}_{\alpha,t}-y_t),\alpha(y_t-\hat{q}_{\alpha,t})
        \right\}
	\right] \quad .
 \end{equation}
Here, gradient boosted machines are used to determine $Q_\alpha$ for $\alpha=0.01,0.05,...,0.95,0.99$, inspired by the winning entry from the 2014 Global Energy Forecasting Competition using the \texttt{R} package \texttt{gbm}~\cite{Landry2016,Ridgeway2014}.

\section{Case Study}
\label{sec:CaseStudy}

The performance of the proposed trading strategies is evaluated in a case study using historic data from the GB power system in the UK. There are two coupled auctions operated by APX and N2EX (Nordpool) which clear at the same price for each hour of the next day. The balancing market comprises half-hour settlement periods and is operated by the System Operator (SO) and Elexon. Trading in the intraday market can take place up until gate closure one hour before each settlement period begins, though participation in this markets is not considered here. Following the end of each settlement period, the single balancing price is calculated based on actions taken by the SO. This price is the volume-weighted average of the most expensive 50MWh of balancing actions taken relating to that period.

Electricity market data are available from Elexon~\cite{elexonPortal}, who operate the data service for the GB balancing mechanism. The data we utilise in this study are day-ahead and balancing prices, plus day-ahead forecasts of load, national wind and solar generation, and generation margin at peak demand. Half-hour resolution wind power and day-ahead power forecasts for five UK wind farm are provided by an anonymous GB wind farm operator and aggregated, since imbalances are calculated on an aggregate basis.

The period 06/11/2015 to 06/05/2016 is used in this case study covering the first six months following the switch from a dual- to single-price balancing mechanism. Due to the limited volume of data, all analysis is performed on a \textit{hold-out} basis where a portion of the data are held-out and used for testing while models are fit to the remaining data.

Offer strategies have been implemented with benchmarks based on perfect and simple forecasts to demonstrate the relative value and limitations of each method.

%
%{\def\arraystretch{1.5}
%\begin{table}
%	\begin{tabular}[t]{c|L{10cm}}
%		Strategy &  Description \\ \hline
%		1.1		& Simple strategy with  perfect forecasts of generation, i.e. $\pi^C$ is received for
%					all generated energy, imbalance costs are zero. \\
%		1.2 	& Simple with day-ahead forecasts issues prior to $t=10:30$am gate closure of the UK
%					day-ahead double-blind auction for the next day's power. \\
%		2.1 	& Perfect categorical forecast of system length, either $0$ or $E_\text{max}$ is bid
%					depending on whether the incentive is to be long or short. \\
%		2.2 	& Categorical estimate of system length based on historic outcomes for each
%					settlement period. \\
%		2.3 	& Categorical forecast of system length based on system information available at time $t$. \\
%		3.1a 	& Probabilistic estimate of system length equal to historic proportions with
%					perfect price forecasts. \\
%		3.2a 	& Probabilistic estimate of system length equal to historic proportions
%					with estimated prices based on mean	for the same settlement period of the same quarter. \\
%		3.3a 	& Probabilistic estimate of system length equal to historic proportions
%					with estimated prices based on mean	for the same settlement period of the same month. \\			
%		3.4a 	& Probabilistic estimate of system length equal to historic proportions
%					with price forecasts. \\
%		3.--b 	& As 3.1a--3.4a but with probabilistic forecasts of system length rather than
%					historic proportion.
%	\end{tabular}
%		\caption{List of bidding strategies employed in case study.}
%		\label{tab:bidStrat}
%\end{table}}

\subsection{System Length Forecast and Evaluation}

The performance of the probabilistic forecast of system length is first evaluated in terms of the Brier score and its decomposition. As a benchmark, the historic proportion of occasion when each settlement period is short is used as a forecast, using the hold-out sample method.

The Brier score is a proper scoring rule for probabilistic forecasts of binary events and is given by
\begin{equation}
%	\text{Brier Score} = \frac{1}{N} \sum_{i=1}^{N} \left( \phi_i - \boldsymbol{1}_{\text{NIV}_i>0} \right)^2
	\text{Brier Score} = \frac{1}{N} \sum_{i=1}^{N} \left( \phi_i - \mathbf{o}_i \right)^2
\end{equation}
where the observation $\mathbf{o}_i =1$ if $\text{NIV}_i>0$, and 0 otherwise~\cite{Brier1950}. The Brier score rewards both reliability and confidence. The best score achievable is 0 if the either 0 or 1 is correctly forecast. Confident forecasts, i.e. those close to 0 or 1, are rewarded with a lower Brier score than cautions perditions, i.e. close to 0.5, if they are correct, and more heavily penalised if they are wrong side of 0.5.

The Brier score can be decomposed into reliability, resolution and uncertainty~\cite{Murphy1973}. Reliability is a measure of how close the forecast probabilities are to the proportion of positive outcomes, resolution is a measure of how much the forecast probabilities vary from the climatic average, and uncertainty measures the inherent uncertainty of the event being forecast. Mathematically these are given by
\begin{eqnarray}
	\text{Reliability} &=& \frac{1}{N}\sum\limits _{k=1}^{K}{n_{k}(\phi_{k}-\bar{\mathbf{o}}_k)}^{2} \quad ,\\
   \text{Resolution} &=& \frac{1}{N}\sum\limits _{k=1}^K n_k({\bar{\mathbf{o}}_k-\bar{\mathbf{o}})}^{2} \quad ,\\
   \text{Uncertainty} &=& \bar{\mathbf{o}}\left(1-\bar{\mathbf{o}}\right) \quad ,
\end{eqnarray}
where $N$ is the total number of forecasts issued, $K$ is the number of unique forecasts issued, and $n_k$ is the total number of times the $k^\text{th}$ unique forecast is issued. The terms $\bar{\mathbf{o}}$ and $\bar{\mathbf{o}}_k$ are the mean outcome and the mean outcome conditional on the $k^\text{th}$ unique forecast being issued, respectively. Here, forecasts are grouped into 21 forecast bins centred on values from 0 to 1 in increments of 0.05.

A separate model is fit for each settlement period. Forecasts of load, wind generation, and generation margin at peak are included as explanatory variables for all periods, while forecast of solar generation are only used during hours of daylight, specifically periods 12--41. Forecasts are produced out-of-sample for each day of the dataset using models trained on the all other data. The performance of this approach is tabulated in Table~\ref{tab:BrierScores}, along with the performance of the benchmark model.

% \begin{table}[t]
% 	\centering
%     \caption{Briar scores for probabilistic system length forecasts from five-fold cross-validation on training dataset. M indicates that the model included system margin, L indicates load, and WF indicates wind forecast. The uncertainty component of the Brier score is 0.2348 in all cases.}
% 	\begin{tabular}{lrrr} \hline
% 	Model & Brier Score & Reliability & Resolution  \\ \hline
% 	M & 0.2372 & 0.0043 & 0.0020 \\
%     L & 0.2357& \textbf{0.0049} & 0.0040 \\
% 	WF & 0.2351 & 0.0030 & 0.0027\\
%     M + L & 0.2343 & 0.0043 &  0.0049\\
%     M + WF & 0.2350 & 0.0040 & 0.0038 \\
%     \textbf{L + WF} & 0.2329 & 0.0046 & 0.0066 \\
%     M + L + WF & \textbf{0.2319} & 0.0042 & \textbf{0.0072} \\ \hline
% %    M + L + WF & 0.2319 & 0.0042 &  0.0072\\
% 	\end{tabular}
% 	\label{tab:SLcrossval}
% \end{table}

%The logistic model fit using load, generation margin, wind and solar forecasts during day-light hours, and the same without solar during hours of darkness.

\begin{table}[t]
	\centering
	\caption{Briar scores for probabilistic system length forecasts. The uncertainty component of the Brier score is 0.2337 in both cases.}
	\begin{tabular}{lcccc} \hline
	Method & Brier Score & Reliability & Resolution  \\ \hline
	Empirical Proportions & 0.2318 & 0.0001   & 0.0029 \\ %& \multirow{2}{2cm}{0.2301} \\
	Logistic Regression & 0.2265 & 0.0030    &  0.0102 \\ \hline
	\end{tabular}
	\label{tab:BrierScores}
\end{table}

% Reliability diagrams illustrate the calibration of a probabilistic
% forecast. For example, for a perfectly calibrated forecast, if an outcome
% is forecast to be observed with probability $p\%$, it should be observed in
% $p\%$ of trials. Reliability diagrams for forecasts of system length based on both empirical proportions and the logistic regression model are shown in Figure~\ref{fig:relDiag}.
% \begin{figure}
% \centering
% 	\epsfxsize 0.85\columnwidth
% 	\mbox{\epsffile{HistoricRel.eps}}
% 	\epsfxsize 0.85\columnwidth
% 	\mbox{ \epsffile{ForecastRel.eps}}
% 	\caption{Reliability diagrams for system length forecasts (1=short) with 95\% confidence intervals. Top: empirical proportions, Bottom: logistic regression.}
% 	\label{fig:relDiag}
% \end{figure}

\begin{figure}
\centering
	\epsfxsize \columnwidth
	\mbox{\epsffile{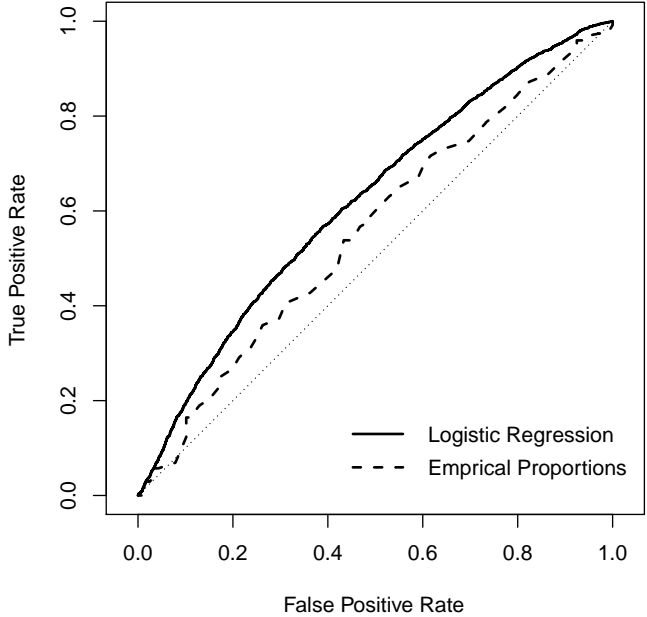}}
	\caption{Relative operator characteristic curves for system length forecasts. The diagonal line, False Positive Rate = True Positive Rate illustrates the performance of a random forecast, e.g. a random forecast of 70\% would be expected to correctly predict 70\% of all positive outcomes, and falsely predict that 70\% of negative outcomes would be positive.}
    %The `area under the curve' for the logistic and empirical forecasts are 0.62 and 0.56, respectively.}
	\label{fig:roc}
\end{figure}

The performance of binary forecasts such as these can also be evaluated by examining their relative operating characteristic (ROC) curves~\cite{Spackman1989,Fawcett2006}. ROC curves depict the trade-off between  true-positive and false-positive forecasts across the full range of predicted probabilities. Loosely, a more skilful forecast method is that with a higher true-positive and lower false-negative rate than the competing method. ROC curves for system length forecasts are presented in Figure~\ref{fig:roc}, which illustrates that forecasts produced by logistic regression consistently outperform the benchmark.

\subsection{Price Forecast Evaluation}

The exogenous variables available for price forecasting are the same as those used in the logistic regression for system length forecasting, namely day-ahead forecasts of load, wind, solar, and generation margin. Data are grouped into three day-types: weekdays, weekends, and holidays. It should be noted that because the day-ahead market requires offers to be submitted before 11am, balancing prices for times later than this will not be available as input to the forecast of balancing prices for the next day. For this reason, two-step-ahead forecasts of balancing prices for periods after 10am are used, to allow for delays in the 10:00--10:30 and 10:30--11:00 balancing prices becoming available.
 
Models of order (1,1) and (2,1) are most common and account for over 25\% of the models fit. Results are presented in terms of the critical probability, $\Phi = \frac{\pi^\text{C} - \pi^-}{\pi^+ - \pi^-}$, and evaluated in terms of root mean square error (RMSE) and mean absolute error (MSE). These are given by
\begin{equation}
 	\text{RMSE} = \sqrt{\frac{1}{N}\sum_{t,k}^N \left(\Phi_{t+k} - \hat{\Phi}_{t+k|t}\right)^2}
\end{equation}
and
 \begin{equation}
 	\text{MAE} = \frac{1}{N}\sum_{t,k}^N \left|\Phi_{t+k} - \hat{\Phi}_{t+k|t}\right|
\end{equation}
where $\hat{\Phi}_{t+k|t}$ is the forecast of $\Phi_{t+k}$ made at time $t$, and $N$ is the total number of samples.

% The price forecast results are evaluated in terms of root mean square error (RMSE) and mean absolute error (MSE). These are given by
%  \begin{equation}
%  	\text{RMSE} = \sqrt{\frac{1}{N}\sum_{t=1}^N (\pi_t - \hat{\pi}_t)^2}
% \end{equation}
% and
%  \begin{equation}
%  	\text{MAE} = \sqrt{\frac{1}{N}\sum_{t=1}^N |\pi_t - \hat{\pi}_t|}
% \end{equation}
% for forecasts $\hat{\pi}_t$ of $\pi_t$, $t=1,...,N$. The mean price for each settlement period of the same month of provides a climatological benchmark. Results of this and the ARMAX models for $\pi^\text{C},~\pi^+$ and $\pi^-$ are presented in Table~\ref{tab:priceResults}.
% \begin{table}
% 	\centering
%       \caption{Error scores for price forecasts in units of \pounds/MWh.}
% 	\begin{tabular}{lcccccc} \hline
%         	& \multicolumn{3}{c}{ARMAX} & \multicolumn{3}{c}{Climatology} \\
%             Metric & $\pi^\text{C}$ & $\pi^+$&  $\pi^-$& $\pi^\text{C}$ & $\pi^+$& $\pi^-$ \\ \hline
%             RMSE & 4.28 & 19.87 & 8.52 & 11.25 & 27.73 & 13.66 \\
%             MAE & 3.21 & 12.83 & 6.31 & 9.04 & 21.08 & 10.50 \\ \hline
%       \end{tabular}
%       \label{tab:priceResults}
% \end{table}

The mean value of day-ahead and balancing prices from the same month and settlement period is used as a simple benchmark to asses the quality of the ARMAX forecasts. The MAE for the ARMAX and simple methods is 0.23 and 0.45, respectively; and the RMSE is 0.43 and 0.56, respectively.

The ARMAX modelling approach clearly outperforms the simple method in terms of both error metrics. Forecasts from both methods will tested though implementation of the bidding strategies described in Sections~\ref{sec:detLength}--\ref{sec:risk_constrained} in order to quantify this improvement in monetary terms.

% The price returned by the day-ahead auction is applicable to two
% settlement periods. Participants who expect to vary their
% power production over an hour-long period submit piece-wise bids
% and are contracted for each balancing settlement period accordingly.

% The single imbalance price is volatile relative to the day-ahead
% price and is bi-modal with the system length.
% This case study uses the prices re-calculated by Elexon for PAR=50
% as part of the p305 consultation since actual data is only available
% from 5 November 2015.

% Climatological forecasts are given by the mean price by settlement
% period and month in the available training data. Advanced forecasts
% are produced using the `similar day' approach combined with exponential
% smoothing. Training data are divided into weekdays, weekends and UK
% holidays and mean prices are calculated for each settlement period.
% Exponential smoothing is then performed on each day-type and settlement
% period combination to produce day-ahead forecasts.

\subsection{Offer Strategy Results}

The revenue generated for the non-risk-constrained strategies described in Sections~\ref{sec:p50}--\ref{sec:probLength} are calculated using the half-hourly metered power from a portfolio of five UK wind farms and the forecasts described above. These results are tabulated in Table~\ref{tab:revenue2} along with results using perfect and simple benchmark forecasts for comparison. Any additional income from subsidies or other incentive schemes is not included, neither are the costs associated with securing access to the transmission system or electricity market membership.

%\begin{table}[t]
%\centering
%	\begin{tabular}{crrrrr}
%	  \hline
%	 Strategy & Sample Revenue & Mean Daily & Mean Weekly & Mean Annual & Per MWh \\ 
%	  \hline
%	1.1 & 196\,373.59 & 1\,442.38 & 10\,096.64 & 525\,025.14 & 44.39 \\ 
%	 1.2 & 195\,364.56 & 1\,434.97 & 10\,044.76 & 522\,327.41 & 44.16 \\ \hline
%	  2.1 & 326\,704.67 & 2\,399.67 & 16\,797.67 & 873\,478.80 & 73.86 \\ 
%	  2.2 & 158\,922.40 & 1\,167.30 & 8\,171.07 & 424\,895.51 & 35.93 \\ 
%	  2.3 & 173\,260.96 & 1\,272.61 & 8\,908.29 & 463\,231.14 & 39.17 \\ \hline
%	  3.1a & 325\,325.32 & 2\,389.54 & 16\,726.75 & 869\,790.98 & 73.54 \\ 
%	  3.1b & 323\,283.53 & 2\,374.54 & 16\,621.77 & 864\,332.03 & 73.08 \\ 
%	  3.2a & 200\,406.56 & 1\,472.00 & 10\,303.99 & 535\,807.72 & 45.30 \\ 
%	  3.2b & 207\,469.21 & 1\,523.87 & 10\,667.12 & 554\,690.44 & 46.90 \\ 
%	  3.3a & 207\,801.44 & 1\,526.32 & 10\,684.21 & 555\,578.69 & 46.98 \\ 
%	  3.3b & 21\,4083.16 & 1\,572.45 & 11\,007.18 & 572\,373.51 & 48.40 \\ 
%	  3.4a & --- & --- & --- & --- & --- \\ 
%	  3.4b & --- & --- & --- & --- & --- \\ 
%	   \hline
%	\end{tabular}
%	\caption{Revenue (\pounds) for Dyffryn Brodyn with trading strategies considered. Sample revenue is that
%	generated for the test period, mean daily, weekly and annual are that normalised
%	to each respective time period. Revenue per-MWh generated in during the test
%	period if listed in the final column.}
%	\label{tab:revenue}
%\end{table}

\begin{table}[t]
\centering
	\caption{Normalised revenue (\pounds/MWh) using different trading strategies. For strategies marked $\ast$ `Forecast Method' refers to the type of price forecast only. The mean day-ahead price during the test period was \pounds34.80/MWh.}
	\begin{tabular}{lrrr}
	  \hline
	  \multirow{2}{1cm}{Strategy}& \multicolumn{3}{c}{Forecast Method} \\
	  & Perfect & Simple & Advanced \\ 
	  \hline
	 Minimise Imbalance		& 34.66 & n/a & 34.82 \\
	 SL Forecast: Deterministic 	& 49.99 & 33.21 & 34.34 \\
	 SL Forecast: Empirical Proportion$^\ast$ 	& 41.75 & 35.94 & 39.27  \\
	 SL Forecast: Logistic$^\ast$	& 41.39 & 36.91 & 39.00 \\
	\hline
% 	\multicolumn{4}{r}{$$ Indicates that columns refer to the type of price forecast only.}
	\end{tabular}
	\label{tab:revenue2}
\end{table}

These results indicate that strategies based on exploiting favourable imbalance prices and a probabilistic forecast of system length can generate more revenue than attempting to minimise imbalance volumes.
The strategy based on a deterministic forecast of system length does not improve on imbalance volume minimisation except in the case of perfect foresight demonstrating the significance of imbalance price asymmetry. All strategies perform best when coupled with advanced rather than simple price forecasts. It is notable that perfect power forecasting does not increase revenue in the imbalance minimisation strategy.

The probabilistic forecast of system length based on logistic regression generates more revenue than that based on empirical proportions using simple price forecasts; however, the converse is true when using the advanced price forecasts, despite the logistic model having superior predictive performance.

% The probabilistic forecast of system length based on empirical proportions generates more revenue than that based on logistic regression, though it is worth highlighting that use of sophisticated price forecasts, which determine threshold probability, are most important. This is likely because only the skill of the system length forecast in the vicinity of the threshold probability will have a significant impact on the final revenue.

Risk constrained strategies are evaluated in terms of revenue, the average size of imbalances, and value at risk (VaR$_\alpha$). The $\alpha\%$ VaR is a threshold value such that the chance of the revenue being below that threshold is $\alpha\%$. Here, it is calculated as the $\alpha$-percentile of the empirical distribution of settlement period revenue. Mean absolute imbalance, given by
\begin{equation}
	\tilde{d} = \text{mean}\left\{|E^\text{C}_{t+k}-E_{t+k}|\right\} \quad ,
\end{equation}
is also reported to compare the size of imbalance leveraged by each strategy.

\begin{figure}[t]
    \centering
	\epsfxsize \columnwidth
	\mbox{\epsffile{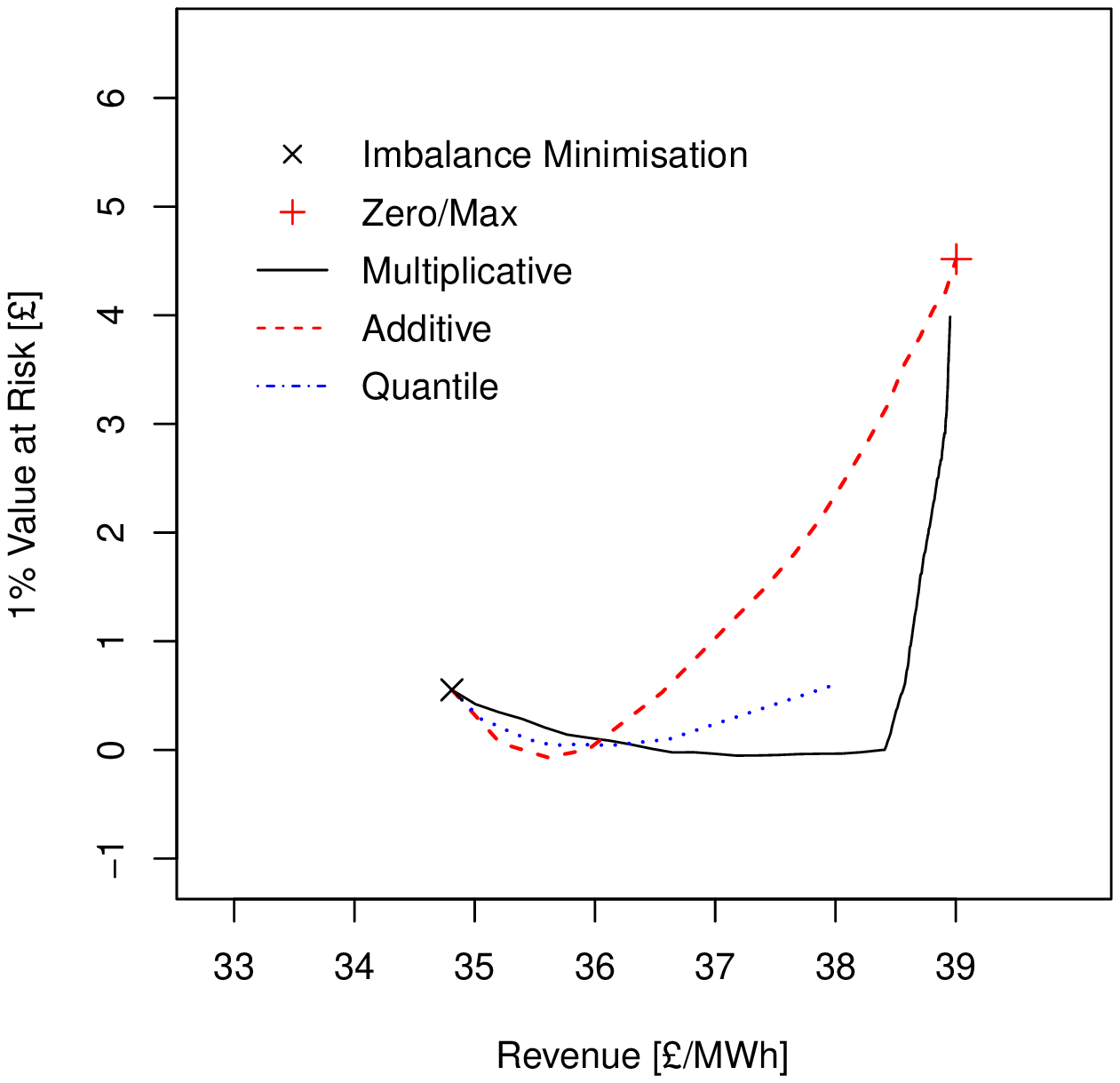}}
	\caption{Plot of revenue vs 1\% VaR
	for the three risk-constrained strategies with parameter values spanning those listed in Table~\ref{tab:DB_risk_results}. Crosses indicate the results from the	revenue-maximising (zero/max) and imbalance minimising strategies tabulated in Table~\ref{tab:revenue2}.}
	\label{fig:DB_rev_var}
\end{figure}

\begin{figure}[t]
\centering
	\epsfxsize \columnwidth
	\mbox{\epsffile{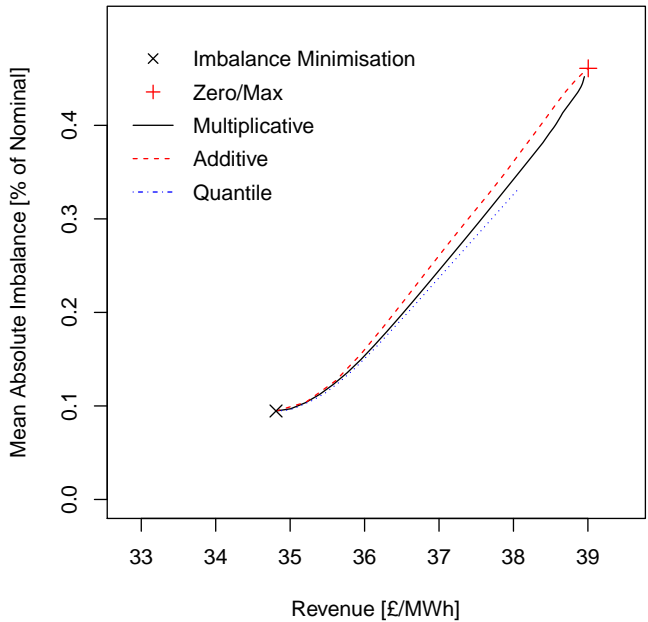}}
	\caption{Plot of revenue vs mean absolute imbalance for the three risk-constrained strategies with parameter values spanning those listed in Table~\ref{tab:DB_risk_results}. Crosses indicate the results from the	revenue-maximising (zero/max) and imbalance minimising strategies tabulated in Table~\ref{tab:revenue2}.}
	\label{fig:DB_rev_imbal}
\end{figure}

Revenue, VaR$_{1\%}$, and mean absolute imbalance are calculated for the three risk-constrained strategies described in Section~\ref{sec:risk_constrained}. The probabilistic forecast of system length from the logistic model is used along with ARMAX forecasts of the prices. Plots of these results are presented in Figures~\ref{fig:DB_rev_var} and~\ref{fig:DB_rev_imbal}. Key results are tabulated in Table~\ref{tab:DB_risk_results}.

All strategies successfully reduce risk and increase revenue when offer volume adjustments are small, but tend towards the high-risk zero/max strategy for larger adjustments. This change in behaviour occurs at the highest revenue for the multiplicative strategy at the point where $\eta=1$, where the offer is either zero or 200\% of expected generation. When $\eta>1$ increasingly extreme short positions are taken while long positions are restricted since offers below zero are not possible.

The additive and quantile strategies, on the other hand, are able to take short positions regardless of expected generation resulting in more frequent and extreme short positions, and therefore, more frequent losses and higher VaR. Since wind power generation is more likely to be close to zero than $E_\text{max}$, the effect described above results in short positions being taken more frequently than long positions. This increases VaR since short positions can result in negative revenue, whereas long positions can only result in reduced revenue, unless the balancing price is negative.

%, particularly expected generation is zero. When $\hat{E}=0$ the multiplicative strategy submits a zero offer, whereas the additive strategy makes a positive offer if is the system is expected to be long with sufficient confidence. This restricts the revenue attainable by the multiplicative strategy but also reduces reduces exposure to possibility of negative returns.

\begin{table}[t]
\centering
	\begin{tabular}{cccc} \hline
		\multicolumn{4}{c}{Additive Adjustment} \\
		$\nu$ & Revenue & VaR$_{1\%}$ & $\tilde{d}$ \\ \hline
        0 & 34.82 & 0.54 & 9\% \\ 
  		0.5 & 37.89 & 2.16 & 35\% \\ 
  		1 & 39.00 & 4.52 & 46\% \\  \hline \hline
		\multicolumn{4}{c}{Multiplicative Adjustment} \\
		$\eta$ & Revenue & VaR$_{1\%}$ & $\tilde{d}$ \\ \hline
      	0 & 34.82 & 0.54 & 9\% \\ 
        0.5 & 36.65 & -0.02 & 21\% \\ 
        1 & 38.40 & -0.00 & 38\% \\ 
        5 & 38.87 & 2.63 & 44\% \\ 
        10 & 38.95 & 3.95 & 45\% \\  \hline \hline
        \multicolumn{4}{c}{Quantile} \\
		$\alpha^\prime$ & Revenue & VaR$_{1\%}$ & $\tilde{d}$ \\ \hline
		0.55 & 34.93 & 0.40 & 10\% \\ 
        0.75 & 35.48 & 0.10 & 11\% \\ 
        0.95 & 36.62 & 0.11 & 20\% \\
        0.99 & 38.05 & 0.62 & 33\% \\ \hline
		Units: & \small \pounds/MWh & \small \pounds & \small \% of $E_\text{max}$ \\ \hline	
	\end{tabular}
	\caption{Performance of the risk-constrained offer strategies. The case $\nu,~\eta=0$ is equivalent to offering a volume equal to the wind power forecast (imbalance minimisation), the additive adjustment strategy with $\nu=1$ is equivalent to offering zero/max.}
	\label{tab:DB_risk_results}
\end{table}

%
%
%\begin{figure}
%\centering
%	\epsfxsize \columnwidth
%	\mbox{\epsffile{revenue_hist.eps}}
%	\caption{Histogram of revenue per settlement period for imbalance volume minimisation (p50), and
%	risk constrained strategies: additive adjustment with $\nu=1$ and multiplicative adjustment with $\eta=1$.}
%	\label{fig:revenue_hist}
%\end{figure}

\section{Concluding Remarks}
\label{sec:Discussion}

Trading strategies for variable generation participating in electricity markets with single-price balancing mechanisms have been proposed and analysed. The problem is formulated as a decision-making problem under uncertainty and solved
relying on a probabilistic forecast of system length, in the first instance to maximise revenue, and in the second to constrain risk. The trading strategies are based on simple analytics using robust and accessible forecasting methods making them adaptable and attractive to many players in the information-rich smart grid paradigm.
%forecasts of multiple variables; however, it has been demonstrated that using `off-the-shelf' methods result are sufficient to add value these strategies.

It is shown that by hedging against penalising imbalance prices, market participants both reduce imbalance charges and profit from increased exposure to favourable balancing prices. In the most extreme example, revenue is increased by over 10\% before considering any subsidy, though this requires the participant to leverage large imbalances that may be considered unacceptable by risk-averse generators, and poor practice by regulators. However, a more conservative risk-constrained approach can increase revenue while simultaneously decreasing risk. While the problem is formulated with the UK electricity market in mind, the principal of positioning oneself favourably in any day-ahead market is applicable to other problems where the cost of correcting that position is reflected in a single price, be that a balancing price or some intraday contract.

Future work should consider extending the problem formulation to include probabilistic price forecasts in order to develop strategies based on the risk associated with specific settlement periods. Furthermore, the limits of the price-taker assumption should be established, and the price-maker scenario studied.

% % % % % % % % % % % % % % % % % % % % % % % % % % % % % % % % % % % % % % % % % % % % % % % % % % % % % % % % % % % % %

\section*{Acknowledgements}
The author acknowledges Elexon and an anonymous wind farm operator for the supply of data. Jethro Browell is supported by the University of Strathclyde's EPSRC Doctoral Prize, grant number EP/M508159/1.

\textbf{Data Statement:} Electricity price and associated data are available to download from Elexon (www.elexon.co.uk) with free registration. Wind power generation data may not be shared to to legal restrictions.

%\vfill
%\pagebreak

%\section*{Bibliography}

\bibliographystyle{IEEEtran}

\bibliography{MyEndNoteLibrary}

\begin{thebibliography}{10}
\providecommand{\url}[1]{#1}
\csname url@samestyle\endcsname
\providecommand{\newblock}{\relax}
\providecommand{\bibinfo}[2]{#2}
\providecommand{\BIBentrySTDinterwordspacing}{\spaceskip=0pt\relax}
\providecommand{\BIBentryALTinterwordstretchfactor}{4}
\providecommand{\BIBentryALTinterwordspacing}{\spaceskip=\fontdimen2\font plus
\BIBentryALTinterwordstretchfactor\fontdimen3\font minus
  \fontdimen4\font\relax}
\providecommand{\BIBforeignlanguage}[2]{{%
\expandafter\ifx\csname l@#1\endcsname\relax
\typeout{** WARNING: IEEEtran.bst: No hyphenation pattern has been}%
\typeout{** loaded for the language `#1'. Using the pattern for}%
\typeout{** the default language instead.}%
\else
\language=\csname l@#1\endcsname
\fi
#2}}
\providecommand{\BIBdecl}{\relax}
\BIBdecl

\bibitem{Edwards2009}
D.~W. Edwards, \emph{Energy Trading and Investing: Trading, Risk Management and
  Structuring Deals in the Energy Market}.\hskip 1em plus 0.5em minus
  0.4em\relax McGraw-Hill, 2009.

\bibitem{Morales2014}
J.~Morales, A.~Conejo, H.~Madsen, P.~Pinson, and M.~Zugno, \emph{Integrating
  Renewable in Electricity Markets}.\hskip 1em plus 0.5em minus 0.4em\relax
  Springer, 2014.

\bibitem{Pei2016}
W.~Pei, Y.~Du, W.~Deng, K.~Sheng, H.~Xiao, and H.~Qu, ``Optimal bidding
  strategy and intramarket mechanism of microgrid aggregator in real-time
  balancing market,'' \emph{IEEE Transactions on Industrial Informatics},
  vol.~12, no.~2, pp. 587--596, April 2016.

\bibitem{Fernandes2016}
C.~Fernandes, P.~Fr{\'i}as, and J.~Reneses, ``Participation of intermittent
  renewable generators in balancing mechanisms: A closer look into the spanish
  market design,'' \emph{Renewable Energy}, vol.~89, pp. 305--316, 2016.

\bibitem{Papakonstantinou2016}
A.~Papakonstantinou and P.~Pinson, ``Information uncertainty in electricity
  markets: Introducing probabilistic offers,'' \emph{IEEE Power Engineering
  Letters}, 2016, in press.

\bibitem{Bathurst2002}
G.~Bathurst, J.~Weatherill, and G.~Strbac, ``Trading wind generation in short
  term energy markets,'' \emph{IEEE Transactions on Power Systems}, vol.~17,
  no.~3, pp. 782--789, 2002.

\bibitem{Bremnes2004}
J.~B. Bremnes, ``Probabilistic wind power forecastsusing local quantile
  regression,'' \emph{Wind Energy}, vol.~7, pp. 47--54, 2004.

\bibitem{Pinson2007}
P.~Pinson, C.~Chevallier, and G.~Kariniotakis, ``Trading wind generation from
  short-term probabilistic forecasts of wind power,'' \emph{IEEE Transaction on
  Power Systems}, vol.~22, no.~3, pp. 1148--1156, 2007.

\bibitem{Bitar2012}
E.~Bitar, R.~Rajagopal, P.~Khargonekar, K.~Poolla, and P.~Varaiya, ``Bringing
  wind energy to market,'' \emph{IEEE Transactions on Power Systems}, vol.~27,
  no.~3, pp. 1225--1235, 2012.

\bibitem{Skajaa2015}
A.~Skajaa, K.~Edlund, and J.~M. Morales, ``Intraday trading of wind energy,''
  \emph{IEEE Transaction on Power Systems}, vol.~30, no.~6, pp. 3181--3189,
  2015.

\bibitem{Zugno2013}
M.~Zugno, J.~M. Morales, P.~Pinson, and H.~Madsen, ``Pool strategy of a
  price-maker wind power producer,'' \emph{IEEE Transactions on Power Systems},
  vol.~28, no.~3, pp. 3440--3450, Aug 2013.

\bibitem{Delikaraoglou2015}
S.~Delikaraoglou, A.~Papakonstantinou, C.~Ordoudis, and P.~Pinson,
  ``Price-maker wind power producer participating in a joint day-ahead and
  real-time market,'' in \emph{12th International Conference on the European
  Energy Market}, May 2015, pp. 1--5.

\bibitem{Weron2014}
R.~Weron, ``Electricity price forecasting: A review of the state-of-the-art
  with a look into the future,'' \emph{International Journal of Forecasting},
  vol.~30, no.~4, pp. 1030 -- 1081, 2014.

\bibitem{Hong2016}
T.~Hong, P.~Pinson, S.~Fan, H.~Zareipour, A.~Troccoli, and R.~J. Hyndman,
  ``Probabilistic energy forecasting: Global energy forecasting competition
  2014 and beyond,'' \emph{International Journal of Forecasting}, 2016, in
  press.

\bibitem{Karakatsani2008}
N.~V. Karakatsani and D.~W. Bunn, ``Forecasting electricity prices: The impact
  of fundamentals and time-varying coefficients,'' \emph{International Journal
  of Forecasting}, vol.~24, no.~4, pp. 764--785, 2008.

\bibitem{Brijs2015}
T.~Brijs, K.~D. Vos, C.~D. Jonghe, and R.~Belmans, ``Statistical analysis of
  negative prices in european balancing markets,'' \emph{Renewable Energy},
  vol.~80, pp. 53--60, 2015.

\bibitem{Ziel2015}
F.~Ziel, R.~Steinert, and S.~Husmann, ``Efficient modeling and forecasting of
  the electricity spot price,'' \emph{Energy Economics}, vol.~47, pp. 98--111,
  2015.

\bibitem{Olsson2008}
M.~Olsson and L.~S{\"o}der, ``Modeling real-time balancing power market prices
  using combined {SARIMA} and {M}arkov processes,'' \emph{IEEE Transaction on
  Power Systems}, vol.~23, pp. 443--450, 2008.

\bibitem{jonsson2014}
T.~J{\'o}nsson, P.~Pinson, H.~A. Nielsen, and H.~Madsen, ``Exponential
  smoothing approaches for prediction in real-time electricity markets,''
  \emph{Energies}, vol.~7, pp. 3710--3732, 2014.

\bibitem{Fang2012}
X.~Fang, S.~Misra, G.~Xue, and D.~Yang, ``Smart grid --- the new and improved
  power grid: A survey,'' \emph{IEEE Communications Surveys Tutorials},
  vol.~14, no.~4, pp. 944--980, 2012.

\bibitem{Hirth2015}
L.~Hirth and I.~Ziegenhagen, ``Balancing power and variable renewables: Three
  links,'' \emph{Renewable and Sustainable Energy Reviews}, vol.~50, pp.
  1035--1051, 2015.

\bibitem{Harris2006}
C.~Harris, \emph{Electricity Markets: Pricing, Structures and Economics}.\hskip
  1em plus 0.5em minus 0.4em\relax Wiley Finance, 2006.

\bibitem{Hyndman2008}
R.~Hyndman and Y.~Khandakar, ``Automatic time series forecasting: The forecast
  package for {R},'' \emph{Journal of Statistical Software}, vol.~26, no.~3,
  2008.

\bibitem{Landry2016}
M.~Landry, T.~P. Erlinger, D.~Patschke, and C.~Varrichio, ``Probabilistic
  gradient boosting machines for {GEFCom2014} wind forecasting,''
  \emph{International Journal of Forecasting}, vol.~32, no.~3, pp. 1061--1066,
  2016.

\bibitem{Ridgeway2014}
\BIBentryALTinterwordspacing
G.~Ridgeway and H.~Southworth, ``gbm: Generalized boosted regression models.
  {R} package version 2.1,'' 2014. [Online]. Available:
  \url{http://CRAN.R-project.org/package=gbm}
\BIBentrySTDinterwordspacing

\bibitem{elexonPortal}
\BIBentryALTinterwordspacing
Elexon. [Online]. Available: \url{www.elexonportal.co.uk}
\BIBentrySTDinterwordspacing

\bibitem{Brier1950}
G.~Brier, ``Verification of forecasts expressed in terms of probability,''
  \emph{Monthly Weather Review}, vol.~78, pp. 1--3, 1950.

\bibitem{Murphy1973}
A.~H. Murphy, ``A new vector partition of the probability score,''
  \emph{Journal of Applied Meteorology}, vol.~12, no.~4, pp. 595--600, 1973.

\bibitem{Spackman1989}
K.~A. Spackman, ``Signal detection theory: Valuable tools for evaluating
  inductive learning,'' in \emph{Proceedings of the Sixth International
  Workshop on Machine Learning}, San Mateo, CA, 1989, pp. 160--163.

\bibitem{Fawcett2006}
T.~Fawcett, ``An introduction to {ROC} analysis,'' \emph{Pattern Recognition
  Letters}, vol.~27, pp. 861--874, 2006.

\end{thebibliography}

%\bibliography{C:/Users/ksb11175/Strathcloud/Personal Folders/Work/Literature/MyEndNoteLibrary}
%\bibliography{D:/Users/ksb11175.DS/GOOGLE\string~1/CDT/Material/MyEndNoteLibrary}

%\vspace{-10mm}
%
%\begin{IEEEbiography}[{\includegraphics[width=1in,height=1.25in,clip,keepaspectratio]{jdowell_sept14.png}}]{Jethro Browell} (M'15) received the M.Phys. degree in Mathematics and Theoretical Physics from the University of St Andrews, UK in 2011 and the Ph.D. degree in Electronic Electrical Engineering from the University of Strathclyde, UK in 2015.
%
%He is currently a Research Associate and holder of an EPSRC Doctoral Prize at the University of Strathclyde where his research interests include energy forecasting and decision-making under uncertainty. Jethro is a member of IEEE PES.
%\end{IEEEbiography}

\end{document}